\def\PNout#1{\sout{#1}}

\def\PNout#1{}
\def\s{$\,$}

\def\vecr{\mathbf{r}}
\def\vecp{\mathbf{p}}

\def\bbsty#1#2#3{{\bf #1} (#3) #2}	
\def\bbstys#1#2#3{{\bf #1}\s (#3)\s #2}	

\documentclass{camera}
\usepackage{graphicx}

\begin{document}

%
\title{Dynamical description \\of heavy-ion collisions at Fermi energies}

%
\author{P. Napolitani$^1$ \and M. Colonna$^2$}

%
\organization{$^1$ IPN, CNRS/IN2P3, Universit\'e Paris-Sud 11, 91406 Orsay cedex, France \\
$^2$ INFN-LNS, Laboratori Nazionali del Sud, 95123 Catania, Italy}

\maketitle

\begin{abstract}
Descriptions of heavy-ion collisions at Fermi energies require to take into account in-medium dissipation and phase-space fluctuations.
The interplay of these correlations with the one-body collective behaviour determines the properties (kinematics and fragment production) and the variety of mechanisms (from fusion to neck formation and multifragmentation) of the exit channel.
Starting from fundamental concepts tested on nuclear matter, we build up a microscopic description which addresses finite systems and applies to experimental observables. 
%
%
%
\end{abstract}

%

%
%


	Heavy-ion collisions at Fermi energies are open systems which require a non--equilibrium dynamical description when the process should be followed from the first instants.
	The evolution of 	the heated system produced in such collisions is determined by the nuclear mean-field
potential, as well as by the explicit action of two-body correlations and fluctuations.
	At large beam energies, when exceeding few hundred MeV per nucleon, those beyond-mean-field contributions become prominent, while the Fermi-energy domain is characterised by the interplay of nucleon-nucleon (N-N) collisions and one-body collective behaviour.
	In order to exploit both mean-field and beyond-mean-field contributions, the one-body description can be extended in terms of BBGKY hierarchy to include N-N collisions and, as their natural corollaries, isoscalar and isovector fluctuations.
	N-N collisions affect flow and stopping~\cite{Lopez2014} while fluctuations produce an ensemble 
of mean field trajectories which reflect in a variety of exit channels and induce fragment formation.
	This representation is the principle of stochastic TDHF which, in the semiclassical context, is analogous to the Boltzmann-Langevin (BL) equation~\cite{Reinhard1992}, written in terms of the one-body distribution function  $f(\vecr,\vecp,t)$ 
\begin{equation}
	\partial_t\,f - \left\{H[f],f\right\} = {\bar{I}[f]}+{\delta I[f]} \;.
\label{eq1}
\end{equation}
The left-hand side gives the Vlasov evolution for $f$ in its own self-consistent mean field; 
the right-hand side contains the average Boltzmann hard two-body collision integral 
$\bar{I}[f]$ and the fluctuating term $\delta I[f]$, both written in terms of the one-body distribution function~\cite{Ayik1990}.
	Numerical solutions of the BL equation have been worked out in different frameworks.
	In the regime of small-amplitude fluctuations, a stochastic definition of the initial states is sufficient, while Fermi energies are mostly related to large-amplitude fluctuations.
	In this latter case, fluctuations can be continuously generated through the collision term: they can be projected on a suited subspace like in the SMF approach~\cite{Colonna1998} or, more efficiently, they can be let develop spontaneously in full phase space from agitating extended portions of the phase space in each single scattering event like in the BLOB approach~\cite{BLOB}.
	The BLOB model constrains the fluctuating term ${\delta I[f]}$ to act on phase-space volumes with the correct occupation variance so that the Pauli blocking is never violated.
	As a consequence, when tested 
for unstable nuclear matter~\cite{BL}, the growth rate of the corresponding (spinodal) unstable modes are connected to the form of the mean-field potential according to the dispersion relation~\cite{Colonna1994}. 
	The propagation of the one-body distribution function is described through the test-particle method; a Skyrme-like (\textit{SKM}$^*$) effective interaction~\cite{Guarnera1996} is employed, defined according to a soft isoscalar equation of state (of compressibility $K\!=\!200$~MeV) and a linear (stiff) density dependence of the potential symmetry energy per nucleon (see~\cite{Napolitani2015} for details). 
%

	In order to have some general examples, we use thereafter the BLOB model to simulate collisions of $^{130}$Xe nuclei. This system is chosen because it recalls widely investigated systems in the region from Sn to Xe~\cite{Esymmbook}; 
in order to rely on simpler entrance-channel properties, the system is chosen symmetric and along $\beta$ stability.
	We investigate firstly central collisions, then we focus on semiperipheral collisions and finally we draw some general prescriptions.

%
%
\begin{figure}[b!]\begin{center}
	\includegraphics[angle=0, width=1\columnwidth]{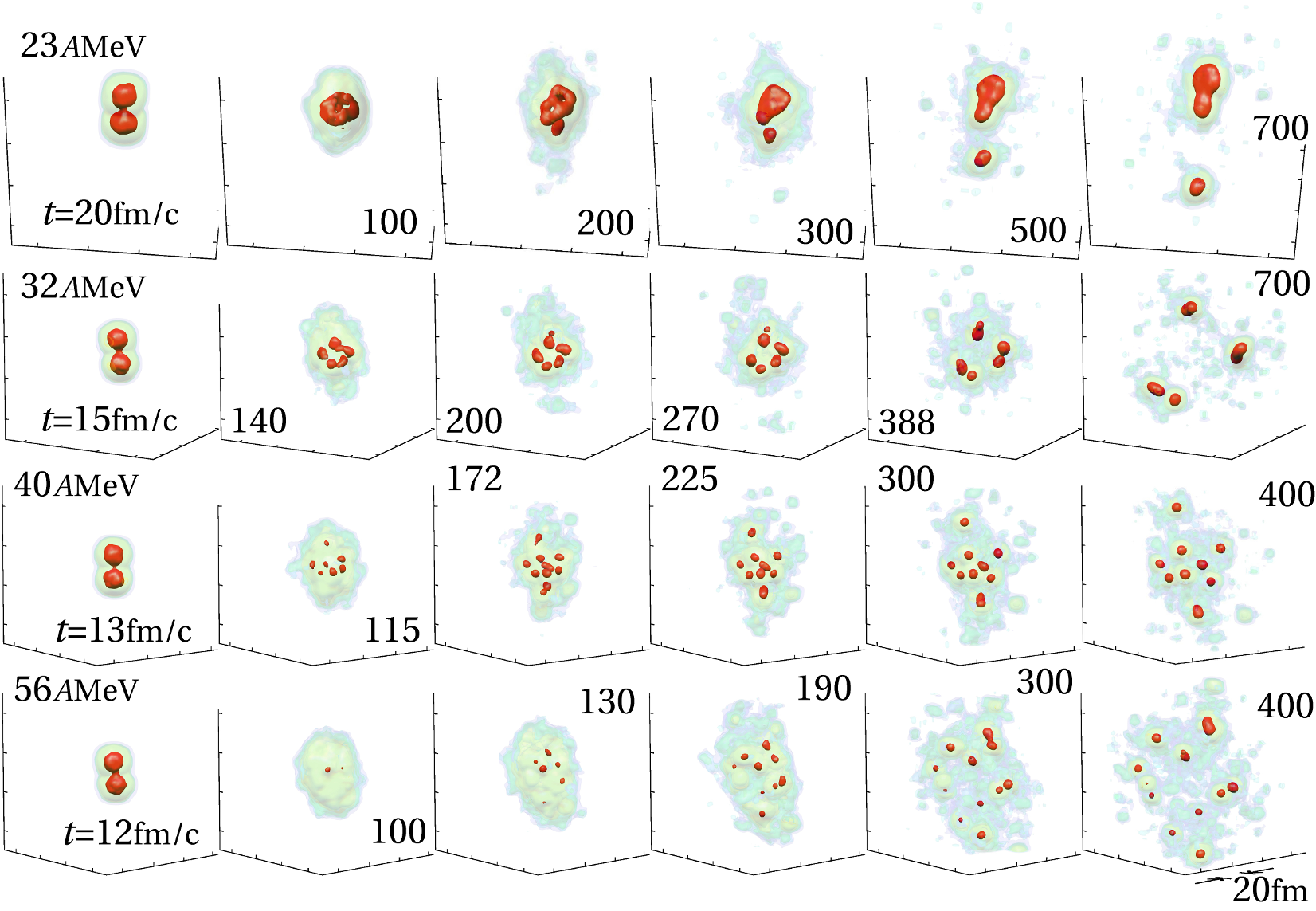}
\end{center}
\caption{
	$^{130}$Xe$+^{130}$Xe, examples of most probable mechanisms leading to IMFs in the exit channel for central impact parameters as a function of the incident energy.
}
\label{fig_order_disorder_map}
\end{figure}
	Head-on collisions are investigated in fig.~\ref{fig_order_disorder_map}, where density distributions are analysed as a function of time for some events related to different incident energies.
	Each event samples the most probable mechanism leading to intermediate-mass fragments (IMFs) in the exit channel for the selected incident energy.
	The event at 32 $A$MeV shows the arising at early times (from 100 to 200 fm/c) of a pattern of several almost-equal-size inhomogeneities in the density landscape.
	These undulations reflect a spinodal behaviour~\cite{Chomaz2004,Borderie2008}, i.e. a condition of mechanical instability where the size of the emerging blobs reflects the leading instability mode of the dispersion relation.
	At later times, if the radial expansion is not sufficient, the equal-size inhomogeneities may reaggregate producing a less symmetric pattern and a small multiplicity of fragments.
	In such chaotic process, the competition between these two antagonist tendencies, the disintegration into several pieces driven by spinodal instability on the one hand, and the action of the attractive nuclear force which tends to bond fragments together, imposes that different exit channels are favoured depending on the bombarding energy.
	The event at 23 $A$MeV shows for instance a highly frustrated fragmentation resulting into an almost complete re-aggregation, so that the final exit channel appears as a very asymmetric binary split.
	It was argued that a similar mechanism also appears in spallation reactions induced by protons and deuterons on heavy nuclei in the 1$A$GeV range~\cite{Napolitani2015}.
	The spinodal signal, imparting fragment-size symmetry, becomes prominent in the event at 40 $A$MeV and fades at larger bombarding energies (i.e. the event at 56 $A$MeV).
	Within this same approach, the analysis of a statistics of exit channels connected to a given set of initial conditions could reveal phase-transition signals in central collisions~\cite{BLOB}.

%
%
\begin{figure}[b!]\begin{center}
	\includegraphics[angle=0, width=1\columnwidth]{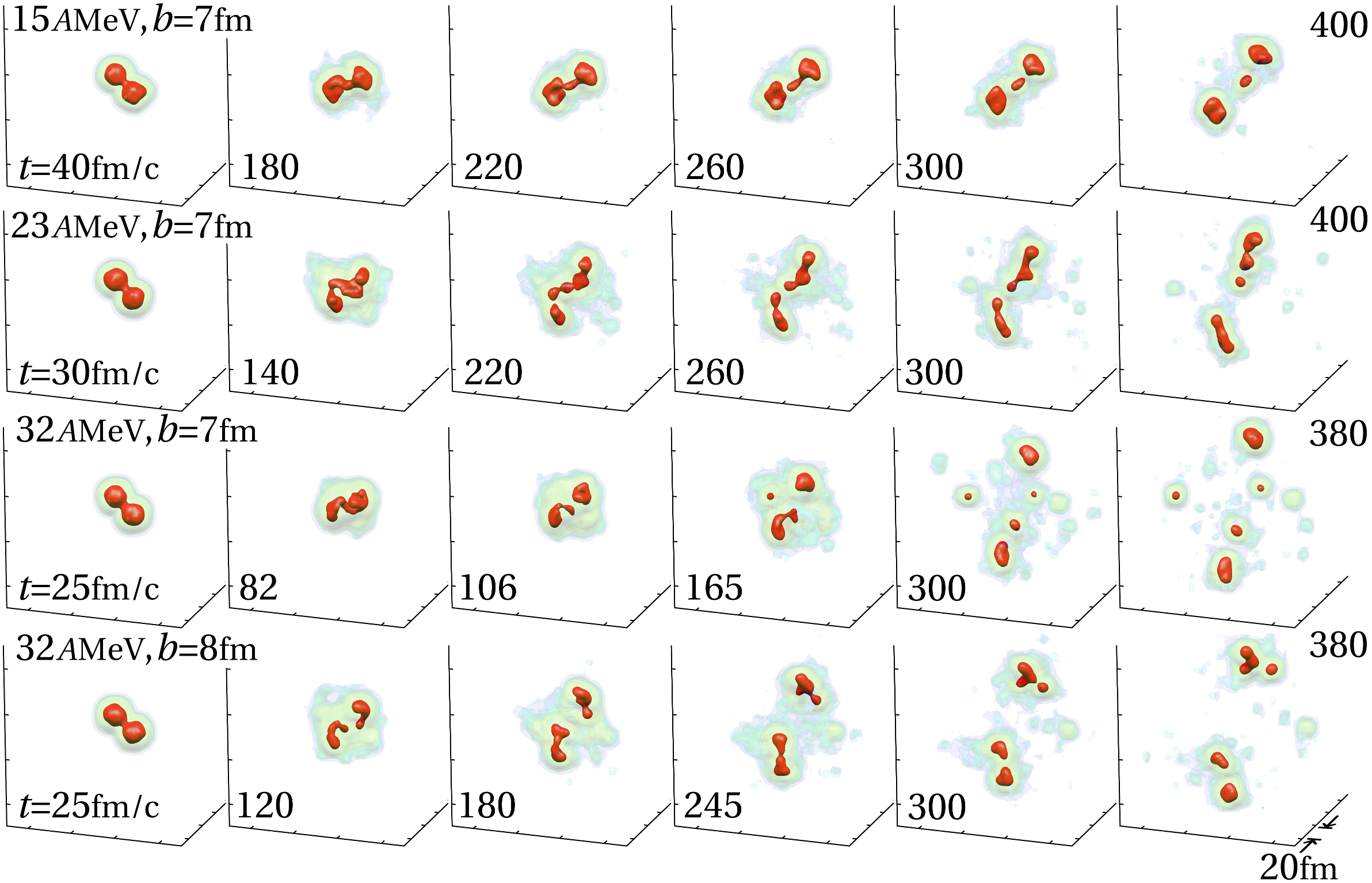}
\end{center}
\caption{
	$^{130}$Xe$+^{130}$Xe, examples of most probable mechanisms leading to IMFs in the exit channel for semiperipheral impact parameters ($b\approx 7$ fm ) as a function of the incident energy and the impact parameter.
}
\label{fig_neck_to_midrapidity_map}
\end{figure}
	Fig.~\ref{fig_neck_to_midrapidity_map} investigates semi-peripheral collisions sampling the most probable mechanisms of IMF production for the selected incident energy.
	At 15 $A$MeV the typical mechanism where IMFs are produced is the formation and separation of a neck region in a rather long process.
	At larger bombarding energies the neck gradually transforms into a diluted midrapidity region where more than one blob can form. 
	At 23 $A$MeV the process is still long and the separation of more than one IMF is too rare.
	At 32 $A$MeV neck fragmentation producing two or more IMFs becomes a favoured mechanism. 
	In less peripheral collisions (i.e. $b=7$ fm) IMFs arise close to the centre of the midrapidity region and they are repelled at large angles with respect to the collision axis, while in more peripheral collisions (i.e. $b=8$ fm) at the same bombarding energy two IMFs tend to form in the proximity of the quasiprojectile (QP) and the quasitarget (QT), respectively; in this case, the IMFs may orbit around the QP or QT and be eventually pulled outside of the collision axis with forward angles.
	These exotic mechanisms were suggested in refs.~\cite{Baran2012_Rizzo2014}. 
%
%
\begin{figure}[b!]\begin{center}
	\includegraphics[angle=0, width=.6\columnwidth]{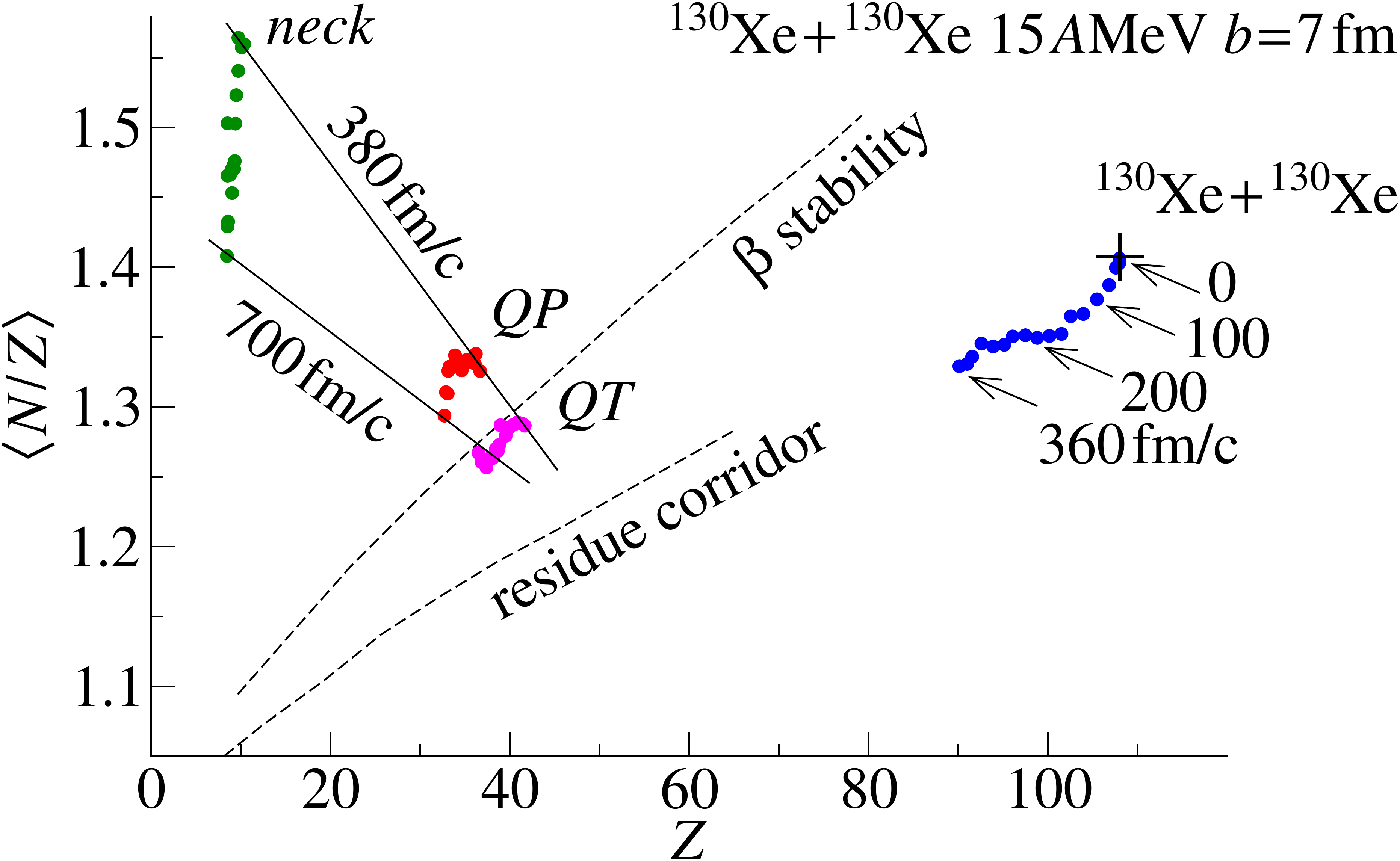}
\end{center}
\caption{
	Neutron enrichment ($N/Z$) of different components of the system $^{130}$Xe$+^{130}$Xe at successive times.
}
\label{fig_neck_NZ}
\end{figure}
	The neck process has been widely investigated especially for its connection to the isospin migration mechanism~\cite{Neck}: this process, driven by density gradients, induces neutron currents from the QP/QT regions towards the diluted midrapidity region where the neck forms.
	As illustrated in fig.~\ref{fig_neck_NZ} for an event at 15 $A$MeV, the consequence is the prominent neutron enrichment of the neck fragment with respect to the QP and QT, which were chosen along $\beta$-stability; in this situation, the hot neck fragment neither succeeds in de-exciting towards the residue corridor, nor it can approach $\beta$-stability.

%
%
\begin{figure}[b!]\begin{center}
	\includegraphics[angle=0, width=.9\columnwidth]{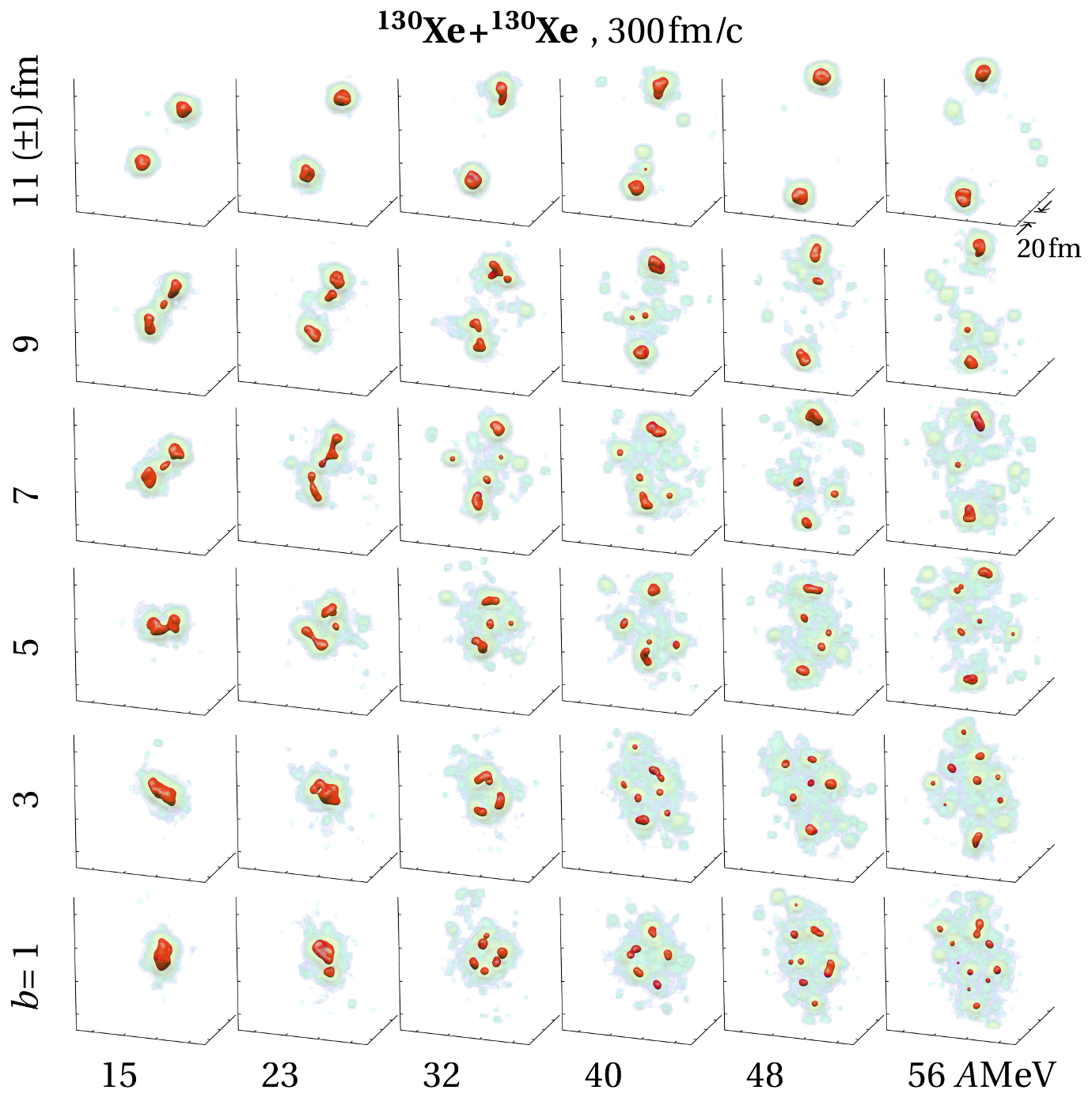}
\end{center}
\caption{
	Examples of most probable configurations with IMFs at 300 fm/c.
}
\label{fig_Fermi_map}
\end{figure}
%
%
%
\begin{figure}[b!]\begin{center}
	\includegraphics[angle=0, width=.9\columnwidth]{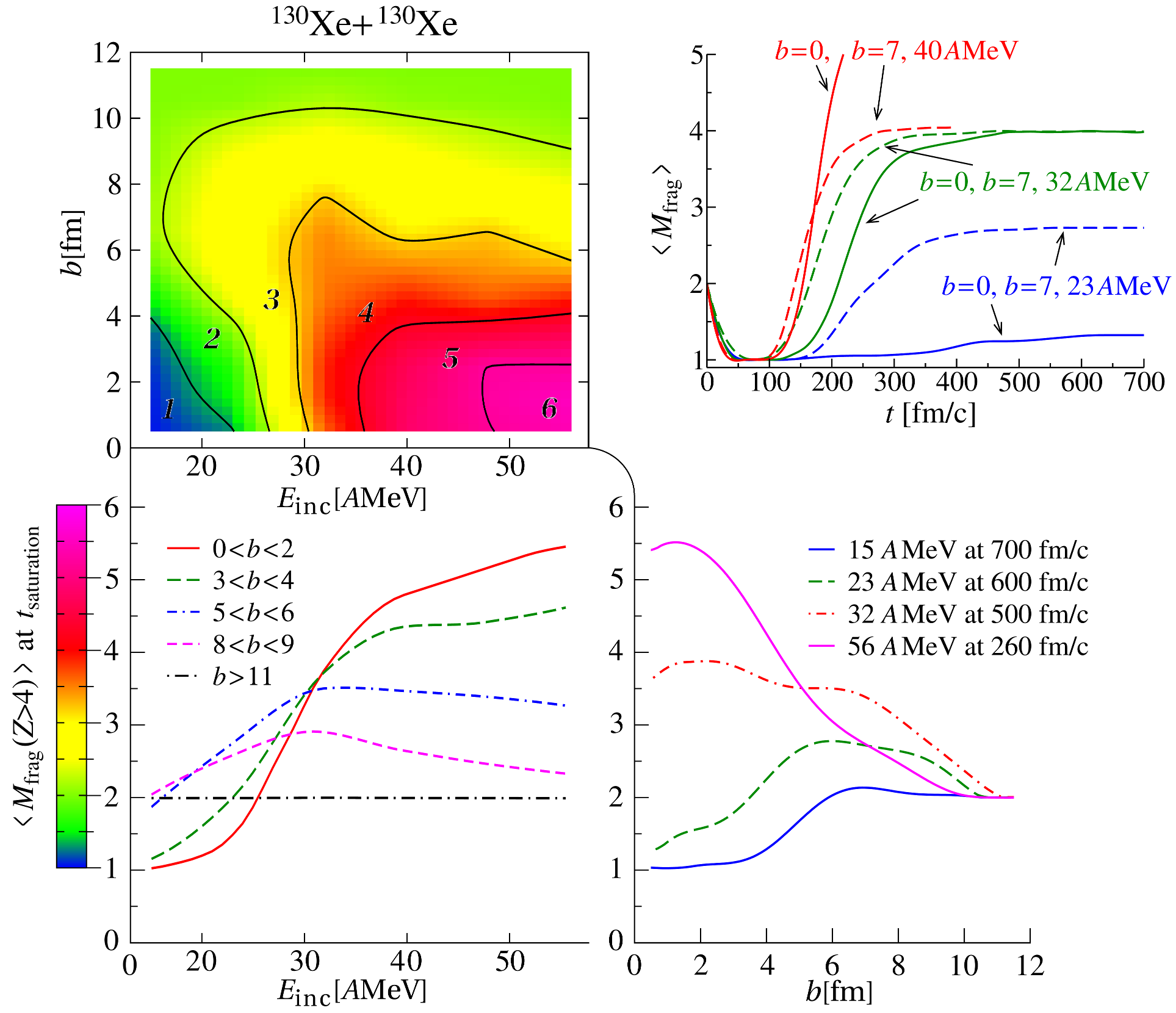}
\end{center}
\caption{
	Top-left and bottom. Mean multiplicity of fragments with $Z\!>\!4$ as a function of impact-parameter $b$ and incident energy $E_{\textrm{inc}}$ for the system $^{130}$Xe$+^{130}$Xe at multiplicity-saturation time; projections on $b$ and $E_{\textrm{inc}}$ are shown in the lower the panels.
	Top-right. Time evolution of the mean multiplicity of fragments with $A>1$ in central and peripheral collisions.
}
\label{fig_Fermiland_bdistr}
\end{figure}
	Samples for the most probable fragment configurations at 300 fm/c for the system $^{130}$Xe$+^{130}$Xe as a function of incident energy and impact parameter are collected in fig.~\ref{fig_Fermi_map}, where all transitions between the different mechanisms discussed above can be followed.
	More quantitatively, the corresponding map of mean multiplicity $\langle M_{\textrm{frag}}(Z\!>\!4) \rangle$ of fragments with $Z>4$ as a function of the impact parameter and incident energy is shown in fig.~\ref{fig_Fermiland_bdistr} (top left), from analysing a statistics of about 3500 events: projections on the two coordinates are also shown in the lower the panels.
	To build the map of fig.~\ref{fig_Fermiland_bdistr} the multiplicity of fragments was extracted at the time when its mean value stops growing.
	The right panel illustrates for some systems that such saturation time can be extracted from analysing the time evolution of the mean multiplicity $\langle M_{\textrm{frag}}\rangle$ of fragments with $A>1$.

	The maps of figs.~\ref{fig_Fermi_map} and \ref{fig_Fermiland_bdistr} indicate in particular the regions where transitions between different mechanisms can arise, and they adapt without significant changes to collisions at Fermi energies of nuclei in the region of Sn and Xe.
	Some of those transitions may be pointed out.
	The fusion cross section fades in favour of asymmetric binary splits above 20 $A$MeV in central collisions and, when approaching 30 $A$MeV, binary splits gradually change into the regular pattern of almost equal-size IMFs, indicating the onset of spinodal multifragmentation.
	Still for central collisions, such symmetric break-up pattern 
persists till around 45 $A$MeV, and further increasing of bombarding energy brings the system outside of the spinodal region.
	Below 20 $A$MeV, along the impact-parameter coordinate $b$, fusion changes into a binary mechanism.
	In the interval of about $5<b<9$ fm, semiperipheral collisions lead to neck formation.
	A transition from ternary channels to events with more than one IMF at mid-rapidity appears when moving to larger bombarding energies (above about 30 $A$MeV) and smaller impact parameters (below about 8 fm).

	In conclusion, we find that a stochastic one-body approach can efficiently apply to dissipative collisions in the Fermi-energy domain, giving hints on the fragment production, the thresholds between different mechanisms, the variety of exit channels and the related isospin properties.
	In particular, we may signal the mechanism of frustrated fragmentation arising from a competition between spinodal instability and mean-field resilience: this process can describe the low-energy threshold of multifragmentation and its possible association to asymmetric splits in two or few fragments.
	We also signal the richness of the process of neck fragmentation, which may result in various unusual patterns to be further investigated in forthcoming theoretical and experimental works.


%
%
%
%

%
%
%
%
\end{document}